# Exciton diffusion in two-dimensional chiral perovskites

Sophia Terres[†], Lucas Scalon[#,‡,¶], Julius Brunner[#,‡], Dominik Horneber[§], Johannes Düreth[§], Shiyu Huang[§], Takashi Taniguchi[∥], Kenji Watanabe[⊥], Ana Flávia Nogueira[¶], Sven Höfling[§], Sebastian Klembt[§], Yana Vaynzof[#,‡] and Alexey Chernikov[†*]

[†] Institute of Applied Physics and Würzburg-Dresden Cluster of Excellence ct.qmat, TUD Dresden University of Technology, 01062 Dresden, Germany

[#] Chair for Emerging Electronic Technologies, TUD Dresden University of Technology, 01187 Dresden, Germany

[‡] Leibniz-Institute for Solid State and Materials Research Dresden, 01069 Dresden, Germany

[¶] Institute of Chemistry, University of Campinas (UNICAMP), 13083-970 Campinas, São Paulo, Brazil

[§] Physikalisches Institut and Würzburg-Dresden Cluster of Excellence ct.qmat, Julius-Maximilians-Universität Würzburg, 97074 Würzburg, Germany

[∥] Research Center for Materials Nanoarchitectonics, National Institute for Material Science, 1-1 Namiki, Tsukuba 305-0044, Japan

[⊥] Research Center for Electronic and Optical Materials, National Institute for Materials Science, 1-1 Namiki, Tsukuba 305-0044, Japan

**ABSTRACT:** Two-dimensional (2D) organic-inorganic hybrid perovskites emerged as a versatile platform for light-emitting and photovoltaic applications due to their unique structural design and chemical flexibility. Their properties depend heavily on both the choice of the inorganic lead halide framework and the surrounding organic layers. Recently, the introduction of chiral cations into 2D perovskites has attracted major interest due to their potential for introducing chirality and tuning the chiro-optical response. Importantly, the optical properties in these materials are dominated by tightly bound excitons that also serve as primary carriers for the energy transport. The mobility of photoinjected excitons is thus important from the perspectives of fundamental material properties and optoelectronic applications, yet remains an open question. Here, we demonstrate exciton propagation in the 2D chiral perovskite methylbenzylammonium lead iodide using transient photoluminescence microscopy and reveal density-dependent transport over more than 100 nanometers at room temperature with diffusion coefficients as high as 2 cm$^2$/s. We observe two distinct regimes of initially rapid diffusive propagation and subsequent localization. Moreover, perovskites with enantiomer pure cations are found to exhibit faster exciton diffusion compared to the racemic mixture, correlated with the impact of the material composition on disorder. Altogether, the observations of efficient exciton diffusion at room temperature highlight the potential of 2D chiral perovskites to merge chiro-optical properties with strong light-matter interaction and efficient energy transport.

Two-dimensional (2D) organic-inorganic hybrid perovskites, originally studied as early as the 1990s,[1–3] emerged as promising platforms for light-emitting[4–6] and photovoltaic applications.[7,8] These semiconducting materials feature an inorganic framework surrounded by organic layers with an exceptional flexibility in their chemical and structural design[9]. The inorganic layers are composed of corner-sharing lead halide octahedra and act as natural quantum wells hosting electronic states forming the conduction and valence bands. The organic cations separate the layers and serve primarily as electronic barriers,[10] while also offering the possibility to integrate a variety of ammonium-based organic cations with different functional groups[11,12] and spatial configurations[13–16]. The ammonium groups bind to the halide of the inorganic framework via hydrogen bonds, while the organic moieties interact with each other through van der Waals and π-stacking interactions.[17]

Recently, chiral organic cations were embedded into perovskites to generate chiro-optical responses,[18,19] highly interesting for polaritonics[20] and optospintronics.[21] Chiral cations were shown to induce structural chirality transfer across the interface between organic and inorganic units of 2D hybrid perovskites by breaking the centrosymmetry of the crystal.[22] This changes the electronic structure, introduces chiral polarization selection rules[18] as well as impacts phase purity and electronic disorder.[23] Moreover, this class of materials inherits the combination of quantum confinement and reduced dielectric screening from the 2D perovskites, giving rise to strong Coulomb interactions between the charge carriers. As a consequence, excitons with binding energies of several 100's meV[1,24,25] form. They represent fundamental electron-hole excitations with strong light-matter coupling[26], dominate the optical response[27] and, most importantly, serve as primarily energy carriers in these systems. Transport of optically injected excitons in chiral 2D perovskites is thus of major interest in the context of both fundamental physics of mobile and chiral many-body states as well as optoelectronic applications. In contrast to more conventional achiral

2D perovskites[28,29], however, the exciton propagation in chiral compounds, the underlying mechanisms, and the relationship to the structure, remain unexplored so far.

Here, we study exciton propagation in 2D chiral perovskites via transient photoluminescence microscopy. We observe linear and non-linear transport in samples with enantiomer pure cations and in the racemic mixtures of both cations. The excitons are found to exhibit initially rapid diffusive transport over more than 100 nanometers at room temperature followed by localization at later times. In addition, excitons propagate faster for all studied densities in perovskites with enantiomer pure cations compared to their racemic mixture counterpart. These findings correlate with different energy scales of disorder determined by photothermal deflection and hyperspatial spectroscopy.

RESULTS

For this study, we used thick-layer samples exfoliated from chiral 2D methylbenzylammonium lead iodide crystals (MBA$_2$PbI$_4$). The crystals were synthesized under nitrogen atmosphere by dissolving lead (II) oxide (PbO) in hydrogen iodide (HI), followed by dropwise addition of R-, S- or rac-methylbenzylamine.[23] Structures of the enantiomer pure chiral perovskite as well the racemic mixture are schematically illustrated in Fig. 1 (a). Enantiomer pure chiral cations induce the formation of asymmetric hydrogen bonds, creating symmetry-breaking distortions in the inorganic framework.[13,22,30] Chemical structures of the cations producing crystallographic right- (R-MBA) and left-handedness (S-MBA) are displayed in Fig. 1 (b). In the racemic mixture, consisting of equal parts S-MBA and R-MBA cations, the hydrogen bonds create symmetric tilting distortions, thus retaining centrosymmetry of the crystal.[17,22] The synthesized bulk crystals were micromechanically exfoliated, transferred onto SiO$_2$/Si substrates using a polymer stamp, and encapsulated between layers of hexagonal boron nitride (h-BN) for environmental protection.[31] The resulting samples comprised of perovskite crystals of a few 100's nm thickness with 10's of nanometers thick h-BN layers were placed in a microscopy cryostat for optical measurements under reduced pressure ($< 10^{-4}$ mbar). We used a continuous wave laser with the photon energy of 3.06 eV for excitation in photoluminescence (PL) mapping and an 80 MHz, 140 fs pulsed Ti:sapphire laser with a photon energy of 3.02 eV for time- and spatially-resolved PL measurements. They were performed at room temperature and the incident laser beams were focused onto the sample by a 60x microscope objective resulting in spot sizes of 0.5 µm. The PL signal was then dispersed using a grating or deflected by a mirror to obtain spectrally and spatially resolved responses, respectively. A CCD camera was used to record time-integrated signals and a streak camera was employed to monitor the time-resolved expansion of the emission area by imaging the luminescence cross-section along the x-coordinate. Additional details regarding sample preparation, cryogenic circularly polarized photoluminescence measurements and photo-thermal deflection spectroscopy are given in the Supplementary Information (SI).

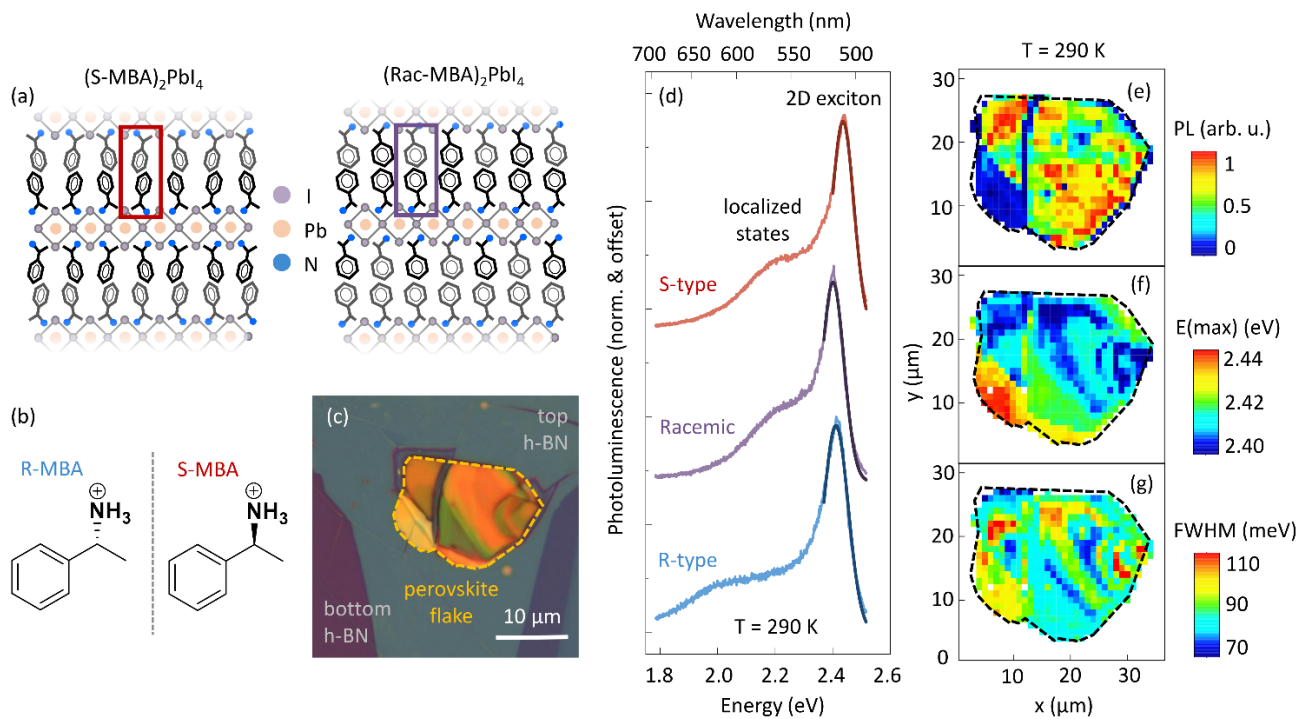

**Figure 1.** (a) Schematic illustration of the layered structure of the (n= 1) chiral perovskite MBA$_2$PbI$_4$, consisting of alternating organic (methylbenzylammonium, MBA) and inorganic layers (PbI$_4$). Left panel: structure of the perovskite with enantiomer pure cations (S-MBA); right panel: perovskite structure with racemic mixture (Rac-MBA). (b) Chemical structure of the enantiomers R-MBA and S-MBA. (c) Optical micrograph of a typical thick-layer sample, encapsulated between 10's of nm thick layers of h-BN. (d) Representative room temperature PL spectra of both S/R-type samples and the racemic mixture at a moderately high excitation fluence of 0.7 µJ/cm$^2$ using 140 fs pulses with photon energy of 3.02 eV. Gaussian fits to the high energy flank of the peaks are indicated by solid lines on top of the data. Intrinsic 2D exciton PL and that attributed to localized states and 1D phase incursions are indicated. (e) Maps of the room temperature PL intensity, (f) PL peak energy and (g) linewidth of an R-type sample with 1 µm step size, measured under continuous-wave 3.06 eV excitation.

**Hyperspectral mapping of chiral 2D perovskites.** A micrograph of a studied thick-layer R-type sample encapsulated between two layers of h-BN is shown in Fig. 1 (c). Typical room temperature PL spectra of all three sample types are presented in Fig. 1 (d). They reveal the dominant 2D exciton signature at approximately 2.4 eV with an asymmetric tail on the lower energy side, commonly associated with localized states[32] and 1D phase incursions.[33,34] While the use of the chiral cations enables the transfer of chirality it also leads to steric hindrance for the interaction of the ammonium group with the inorganic core, resulting in the formation of 1D moieties within the 2D perovskite giving rise to broad-band emission at lower energies.[34,35]

The position of the main exciton peak and the lower energy shoulder depend on the sample position with variations of up to a few ten's of meV. We thus perform hyperspectral PL mapping to assess spatial variations of the exciton spectral features on the micrometer scale. Figures 1 (e) - (g) display maps of the PL intensity, peak energy and linewidth exemplary for an R-type sample (see SI for S-type and racemic mixture). Each pixel on the map corresponds to an individual PL spectrum. The values for the different parameters are extracted from Gaussian fits to the high energy flank of the 2D exciton resonance as indicated by solid lines in the respective colors in Fig. 2 (d). The map of the extracted PL intensity reveals spot-to-spot variations on the order of 50 %. They are partially related to fluctuations in material thickness across the exfoliated sample, as observed by the changes of color due to interference effects in the micrograph in Fig. 1 (c). However, PL intensity varies to a smaller degree also in regions of seemingly uniform thickness, - an observation not untypical for a variety of 2D materials[36], including perovskites[37]. Analysis of the map of the PL peak energy shows an energy landscape featuring both homogeneous areas of reasonably flat potentials with deviations of only a few meV over many µm, but also variations between them. We observe local energy shifts of the exciton resonance on the order of 10 to 20 meV and overall shifts of up to 40 meV across the sample. Variations in the total linewidth within one measurement spot are on the same order of magnitude, being in reasonable agreement with the inhomogeneities in the energy distribution on the larger scale. This represents the overall potential landscape for the excitons to propagate in the studied material.

**Exciton diffusion.** To study exciton transport, we employ time- and spatially resolved emission microscopy[38,39], schematically illustrated in Fig. 2 (a). Ultrafast laser pulses with excitation energy densities of $0.05 - 1.2$ µJ/cm$^2$ create a local distribution of excitons and the expansion of the exciton cloud is detected as function of space and time. A typical PL transient of an S-type sample recorded at moderate excitation density of 0.7 µJ/cm$^2$ is presented in Fig. 2 (b). It features a fast decay of the exciton population over the first several 100's of ps after excitation and slower decay dynamics at later times. To analyze exciton transport across these regimes, we extract the broadening $\sigma$ of the PL emission profiles from a Gaussian fit of the form $\exp[-x^2/2\sigma(t)^2]$. The resulting mean squared displacement $\Delta\sigma^2$ as a function of time is presented in Fig. 2 (c) and exhibits two distinct propagation regimes.[40] At early times, we find linear increase in the exciton spatial distribution, characteristic for diffusive transport[38]. The diffusion coefficient $D$ is extracted from the slope of the variance according to $\Delta\sigma^2 = \sigma^2(t) - \sigma^2(0) = 2Dt$.[38] The obtained value of 0.7 cm$^2$/s is very similar to findings in (PEA)$_2$PbI$_4$[40,41], in which the organic cation is a constitutional isomer of MBA, known to exhibit comparatively

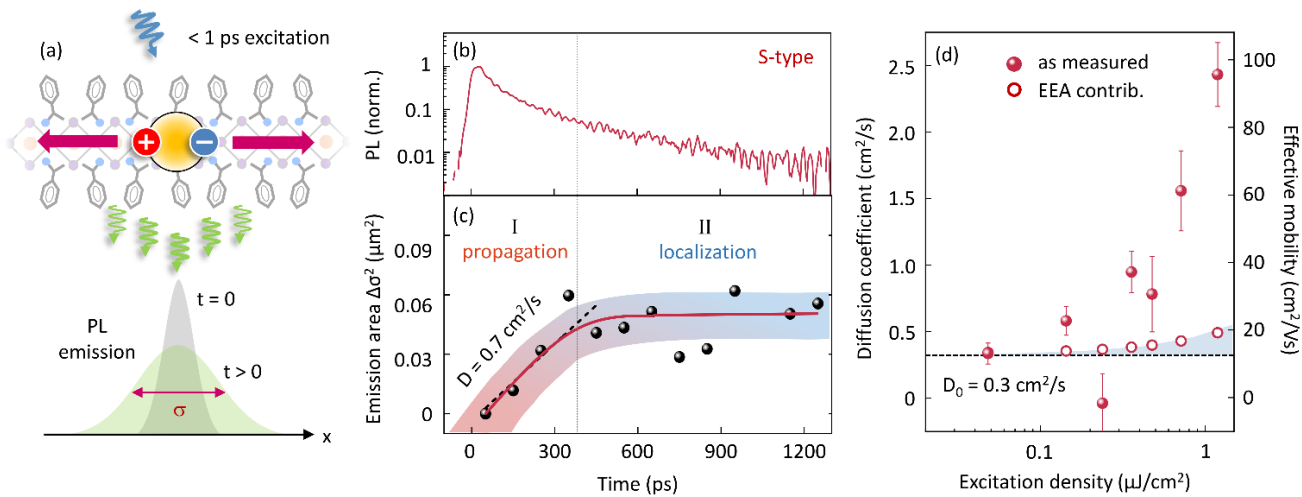

**Figure 2** (a) Schematic illustration of propagating excitons in the chiral perovskite MBA$_2$PbI$_4$ and the resulting emission, monitored via time- and spatially-resolved photoluminescence microscopy. (b) Transient PL in an S-type sample, exhibiting initially rapid decay followed by slower dynamics. (c) Corresponding mean-free displacement as a function of time, demonstrating a linear increase during the first 100's of ps characteristic for diffusive transport (regime I). It saturates at later times indicating localization (regime II). The dotted black vertical line serves as a separation between the two regimes (I & II). The diffusion coefficient D is extracted from the slope of a linear fit (black dashed line). Shaded area with the red line are guides-to-the-eye. (d) Diffusion coefficients as function of excitation energy density demonstrating density-activated behavior. Estimated contribution of exciton-exciton annihilation (EEA) obtained from initial decay rate analysis shows negligible influence of this process. The black dashed line indicates the low-density diffusion coefficient D$_0$.

high diffusion in contrast to other 2D perovskites with different types of cations[28,29]. At later times, broadening of the PL emission reaches saturation, defining a second regime where the absence of spatial expansion indicates localization of excitons as they get trapped in lower-energy sites.[32] This behavior can be found in perovskites with achiral cations depending on background illumination and excitation density.[28,40,42] To test for non-linear processes at elevated fluences[29], we thus performed diffusion measurements as a function of excitation energy density on the S-type sample. The results are shown in Fig. 2 (d) from low densities to the regime where the excitation starts to reduce relative PL intensity due to photo-bleaching. The diffusion coefficients increase linearly with excitation density, reaching values of up to 2.4 cm$^2$/s (corresponding to an effective mobility of 95 cm$^2$/Vs). This result is in stark contrast to fluence-dependent measurements performed on (PEA)$_2$PbI$_4$, where diffusion coefficients remain nearly constant over more than three orders of magnitude in excitation density.[40,41] In general, density-activated diffusion can be indicative of the emergence of non-linear processes, such as exciton-exciton annihilation (EEA) leading to a strong increase in the observed diffusion coefficients and reduction of exciton lifetimes.[43,44]

We estimate the EEA coefficient by analyzing the initial decay rate of the PL after the excitation (see SI) and determine the resulting effective diffusion coefficients according to $D_{eff} = D_0 + \frac{R_A \cdot n_0 \cdot w_0^2}{16}$. Here, $R_A$ is the EAA coefficient, $D_0$ is the diffusion coefficient in the low density limit and $w_0$ the width of the laser profile derived from the FWHM ($w_0 = \frac{FWHM}{2\sqrt{ln2}}$).[43] The determined $R_A$ of 8x10$^{-3}$ cm$^2$/s would correspond to an increase of the effective diffusion coefficients of only 0.2 cm$^2$/s over the studied density range, therefore only accounting for a negligible part of the observed increase. Similar analysis of excitation density dependence in R- and Rac-type samples yield comparable results (see SI). This means that exciton-exciton interactions only marginally contribute and the non-linear diffusivity has a different origin in the studied samples. A possible alternative is the gradual filling of the disorder landscape with rising excitation density towards the excitonic mobility edge leading to increasingly efficient exciton transport.

**Exciton diffusion across different chiralities**. To investigate how structural changes to the inorganic framework affect exciton transport properties, we monitor propagation in enantiomer-pure perovskites (R/S-types) and the racemic mixture (Rac-type) at two distinct excitation fluences. The diffusion coefficient measurements recorded at low density (0.14 µJ/cm$^2$) are summarized in Fig. 3 (a) and the ones measured at higher excitation energy fluence (0.7 cm$^2$/s) are shown in Fig. 3 (b). The box plots display the interquartile range with upper and lower quartiles being the 25$^{th}$ and 75$^{th}$ percentiles. The horizontal line in the center of the box indicates the median diffusion coefficient value for each sample type. The individual measurements are presented together with the box plots and reach values up to 0.9 cm$^2$/s even at low excitation densities, while also show a considerable spread in all studied cases. We note, that the observation of substantial fluctuations in the diffusion coefficient is not unusual in 2D hybrid perovskites.[41] It is also in line with the energy fluctuations of the exciton peak determined by hyperspatial microscopy (see Fig. 1 (f)). Nevertheless, median transport coefficients show that the propagation of excitons in samples with enantiomer pure cations is faster compared to racemic mixtures. This applies for both low and high excitation fluence with S-type samples demonstrating the highest overall diffusion coefficients.

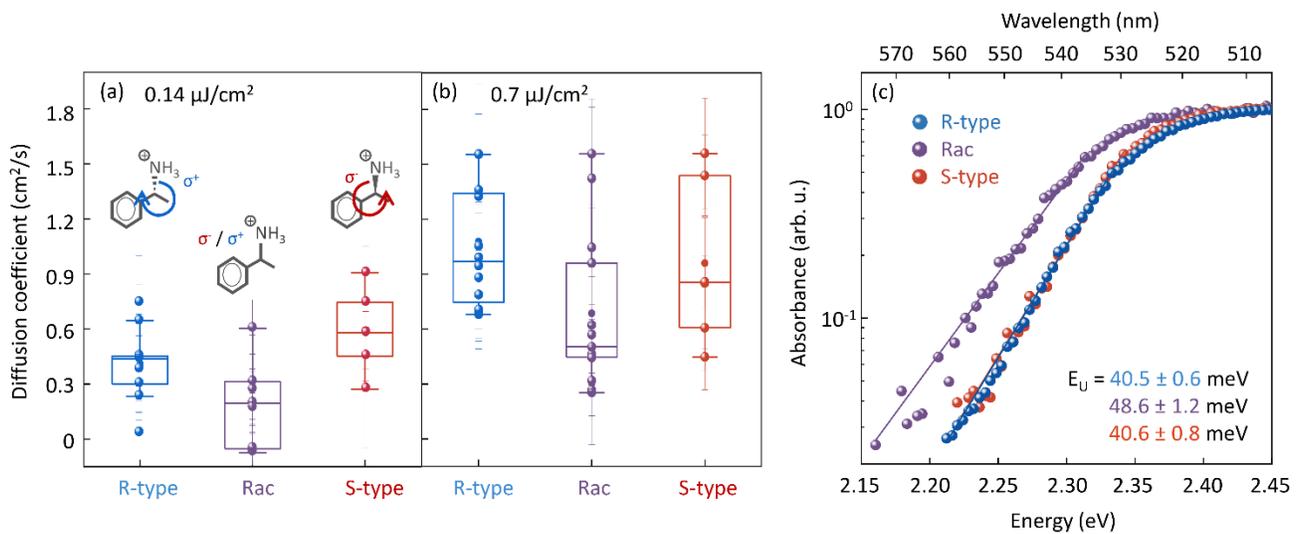

**Figure 3** (a) Summary of measured exciton diffusion coefficients recorded at low excitation density of 0.14 µJ/cm$^2$ for all three sample types (R/Rac/S). The box plots display the interquartile range with the upper and lower quartiles representing the 25$^{th}$ and 75$^{th}$ percentiles, respectively. The line within each box indicates the median value, while the dot in the center represents the average. The box plots are superimposed with the diffusion coefficients of the corresponding individual measurements. (b) Corresponding data for higher excitation fluence of 0.7 µJ/cm$^2$ in analogy to (a). (c) Absorbance spectra from photo-thermal deflection measurements using thicker crystals. The disorder is characterized by the Urbach energies $E_U$ extracted from exponential fits (solid lines) and is correlated with the average diffusion coefficients.

This observation points to structure-related differences in the exciton energy landscape, considering that chiral cations can introduce distortions to the inorganic framework. One of the metrics that is particularly sensitive to that is the detection of Urbach energies $E_U$ used to quantify disorder in the crystal lattice. The parameter contains contributions from both static and dynamic disorder and usually ranges from a few to several tens of meV in chiral 2D perovskites.[45] In amorphous semiconductors the Urbach energy is mostly dominated by static disorder caused by variations in bond length and bond angle.[46] In perovskites such as $MAPbI_3$, however, a substantial contribution can also arise from the dynamic component due to the cage vibrations of the inorganic framework.[45,47]

To determine the disorder parameter $E_U$ in the studied racemic mixtures compared to enantiomer pure crystals we thus employ photo-thermal deflection spectroscopy using thick, large area samples. The resulting absorbance spectra for all three sample types are presented in Fig. 3 (c). The spectra exhibit an exponential decrease towards lower energies from excitonic tail states caused by defects in the crystal structure and lattice vibrations.[48] From exponential fits to the data, we extract Urbach energies of 40.5 and 40.6 meV for R- and S-type samples and 48.6 meV for the racemic mixture. Overall, the obtained values for the Urbach energies are somewhat larger compared to thin films of $PEA_2PbI_4$ with $E_U = 35$ meV[49] and $MBA_2PbI_4$ with $E_U = 29$ meV.[23] Most importantly, they naturally explain differences in the observed median diffusion coefficients between enantiomer pure 2D chiral perovskites and racemic mixtures.

CONCLUSION

In summary, we have experimentally demonstrated efficient exciton diffusion in single crystals of chiral 2D perovskites with values on the order of 1 $cm^2$/s and the ability to efficiently transport energy over 100's of nm at room temperature. We identified two distinct regimes of initially rapid propagation within the first 0.5 ns and subsequent localization at later times. The studied crystals exhibited areas with homogenous energy distribution on the order of several microns but also 10's of meV energy variations on larger spatial scales in accordance with the determined Urbach tail values. Moreover, excitons are found to propagate faster in perovskites with enantiomer pure cations compared to the racemic mixture of both cations, linked to differences in disorder. Finally, we find a strong fluence dependence of the transport coefficients for all studied samples in contrast to other perovskites with similar, yet achiral cations, with only negligible impact of exciton-exciton annihilation. Altogether, the observation of comparatively fast room temperature exciton transport despite the presence of disorder renders chiral 2D perovskites an interesting platform for chiro-optical devices featuring mobile exciton quasiparticles. Alternatively, localization of excitons could be leveraged towards developing single-photon sources based on this class of chiral materials. Future developments towards either increased excitonic mobility or the creation of exciton traps seem both promising in view of the flexibility in the 2D perovskites' design.

AUTHOR INFORMATION

**Corresponding Author**

* Alexey Chernikov - Institute of Applied Physics and Würzburg-Dresden Cluster of Excellence ct.qmat, TUD Dresden University of Technology, 01062 Dresden, Germany; Email: alexey.chernikov@tu-dresden.de

**Author Contributions**

A.C. and S.T. conceived the experimental idea, together with L.S and Y.V.. S.T encapsulated the samples, performed the hyperspectral mapping and transient photoluminescence microscopy experiments. L.S, A.F.N. and Y.V. synthesized the perovskite crystals. K.W. and T.T. provided h-BN crystals. J.I.B. carried out the photothermal deflection spectroscopy measurements. D.H., J.D., Sh.Hu. Sv.Hö. and S.K. performed circularly polarized photoluminescence measurements. The manuscript was written by S.T. and A.C. with input from all authors.

**Funding Sources**

Deutsche Forschungsgemeinschaft (DFG): SPP2196 (424709454, 424216076); CRC1415 (417590517); GRK 2767; Excellence Cluster EXC 2147: 390858490;
Sächsisches Staatsministerium für Wissenschaft, Kultur und Tourismus (SMWK): R.8003.22
The São Paulo Research Foundation (FAPESP): 2017/11631-2, 2018/21401-7;
Japan Society for the Promotion of Science (JSPS): 21H05233, 23H02052;
Brazil's National Oil, Natural Gas and Biofuels Agency (ANP);
World Premier International Research Center Initiative (WPI)

**Notes**

The authors declare no competing financial interest.

ACKNOWLEDGMENT

Financial support by the Deutsche Forschungsgemeinschaft (DFG) via SPP2196 Priority Program (Project-ID: 424709454), CRC1415 (Project-ID: 417590517, B11) and the Würzburg-Dresden Cluster of Excellence on Complexity and Topology in Quantum Matter ct.qmat (EXC 2147, Project-ID 390858490) is gratefully acknowledged. L.S. thanks the São Paulo Research Foundation (FAPESP), grants number 2020/04406-5 and 2021/12104-1. L.S. and A.F.N. acknowledge the support from FAPESP (grant Numbers 2017/11631-2 and 2018/21401-7), Shell, and the strategic importance of the support given by ANP (Brazil's National Oil, Natural Gas and Biofuels Agency) through the R&D levy regulation. Y.V. thank the DFG for funding in the framework of the Special Priority Program (SPP 2196) project PERFECT PVs (Project-ID: 424216076) and for generous support within the framework of the GRK 2767 (project A7). Part of the work was performed within the frame of the M-ERA.NET project PHANTASTIC (R.8003.22), supported by the SMWK. K.W. and T.T. acknowledge support from the JSPS KAKENHI (Grant Numbers 21H05233 and 23H02052) and World Premier International Research Center Initiative (WPI), MEXT, Japan.

# REFERENCES

(1) Ishihara, T.; Takahashi, J.; Goto, T. Exciton State in Two-Dimensional Perovskite Semiconductor (C10H21NH3)2PbI4. *Solid State Commun.* **1989**, *69* (9), 933–936. https://doi.org/10.1016/0038-1098(89)90935-6.

(2) Mitzi, D. B.; Feild, C. A.; Harrison, W. T. A.; Guloy, A. M. Conducting Tin Halides with a Layered Organic-Based Perovskite Structure. *Nature* **1994**, *369* (6480), 467–469. https://doi.org/10.1038/369467a0.

(3) Muljarov, E. A.; Tikhodeev, S. G.; Gippius, N. A.; Ishihara, T. Excitons in Self-Organized Semiconductor/Insulator Superlattices: PbI-Based Perovskite Compounds. *Phys. Rev. B* **1995**, *51* (20), 14370–14378. https://doi.org/10.1103/PhysRevB.51.14370.

(4) Kumar, S.; Jagielski, J.; Yakunin, S.; Rice, P.; Chiu, Y.-C.; Wang, M.; Nedelcu, G.; Kim, Y.; Lin, S.; Santos, E. J. G.; Kovalenko, M. V.; Shih, C.-J. Efficient Blue Electroluminescence Using Quantum-Confined Two-Dimensional Perovskites. *ACS Nano* **2016**, *10* (10), 9720–9729. https://doi.org/10.1021/acsnano.6b05775.

(5) Tsai, H.; Nie, W.; Blancon, J.; Stoumpos, C. C.; Soe, C. M. M.; Yoo, J.; Crochet, J.; Tretiak, S.; Even, J.; Sadhanala, A.; Azzellino, G.; Brenes, R.; Ajayan, P. M.; Bulović, V.; Stranks, S. D.; Friend, R. H.; Kanatzidis, M. G.; Mohite, A. D. Stable Light-Emitting Diodes Using Phase-Pure Ruddlesden–Popper Layered Perovskites. *Adv. Mater.* **2018**, *30* (6), 1704217. https://doi.org/10.1002/adma.201704217.

(6) Wang, N.; Cheng, L.; Ge, R.; Zhang, S.; Miao, Y.; Zou, W.; Yi, C.; Sun, Y.; Cao, Y.; Yang, R.; Wei, Y.; Guo, Q.; Ke, Y.; Yu, M.; Jin, Y.; Liu, Y.; Ding, Q.; Di, D.; Yang, L.; Xing, G.; Tian, H.; Jin, C.; Gao, F.; Friend, R. H.; Wang, J.; Huang, W. Perovskite Light-Emitting Diodes Based on Solution-Processed Self-Organized Multiple Quantum Wells. *Nat. Photonics* **2016**, *10* (11), 699–704. https://doi.org/10.1038/nphoton.2016.185.

(7) Grancini, G.; Nazeeruddin, M. K. Dimensional Tailoring of Hybrid Perovskites for Photovoltaics. *Nat. Rev. Mater.* **2018**, *4* (1), 4–22. https://doi.org/10.1038/s41578-018-0065-0.

(8) Tsai, H.; Nie, W.; Blancon, J.-C.; Stoumpos, C. C.; Asadpour, R.; Harutyunyan, B.; Neukirch, A. J.; Verduzco, R.; Crochet, J. J.; Tretiak, S.; Pedesseau, L.; Even, J.; Alam, M. A.; Gupta, G.; Lou, J.; Ajayan, P. M.; Bedzyk, M. J.; Kanatzidis, M. G.; Mohite, A. D. High-Efficiency Two-Dimensional Ruddlesden–Popper Perovskite Solar Cells. *Nature* **2016**, *536* (7616), 312–316. https://doi.org/10.1038/nature18306.

(9) Mitzi, D. B. Synthesis, Structure, and Properties of Organic-Inorganic Perovskites and Related Materials. In *Progress in Inorganic Chemistry*; Karlin, K. D., Ed.; Wiley, 1999; Vol. 48, pp 1–121. https://doi.org/10.1002/9780470166499.ch1.

(10) Ishihara, T.; Takahashi, J.; Goto, T. Optical Properties Due to Electronic Transitions in Two-Dimensional Semiconductors ( C n H 2 n + 1 NH 3 ) 2 PbI 4. *Phys. Rev. B* **1990**, *42* (17), 11099–11107. https://doi.org/10.1103/PhysRevB.42.11099.

(11) Zhang, F.; Kim, D. H.; Lu, H.; Park, J.-S.; Larson, B. W.; Hu, J.; Gao, L.; Xiao, C.; Reid, O. G.; Chen, X.; Zhao, Q.; Ndione, P. F.; Berry, J. J.; You, W.; Walsh, A.; Beard, M. C.; Zhu, K. Enhanced Charge Transport in 2D Perovskites via Fluorination of Organic Cation. *J. Am. Chem. Soc.* **2019**, *141* (14), 5972–5979. https://doi.org/10.1021/jacs.9b00972.

(12) Zhao, X.; Ball, M. L.; Kakekhani, A.; Liu, T.; Rappe, A. M.; Loo, Y.-L. A Charge Transfer Framework That Describes Supramolecular Interactions Governing Structure and Properties of 2D Perovskites. *Nat. Commun.* **2022**, *13* (1), 3970. https://doi.org/10.1038/s41467-022-31567-y.

(13) Ahn, J.; Lee, E.; Tan, J.; Yang, W.; Kim, B.; Moon, J. A New Class of Chiral Semiconductors: Chiral-Organic-Molecule-Incorporating Organic–Inorganic Hybrid Perovskites. *Mater. Horiz.* **2017**, *4* (5), 851–856. https://doi.org/10.1039/C7MH00197E.

(14) Han, C.; Bradford, A. J.; McNulty, J. A.; Zhang, W.; Halasyamani, P. S.; Slawin, A. M. Z.; Morrison, F. D.; Lee, S. L.; Lightfoot, P. Polarity and Ferromagnetism in Two-Dimensional Hybrid Copper Perovskites with Chlorinated Aromatic Spacers. *Chem. Mater.* **2022**, *34* (5), 2458–2467. https://doi.org/10.1021/acs.chemmater.2c00107.

(15) Lin, J.; Chen, D.; Yang, L.; Lin, T.; Liu, Y.; Chao, Y.; Chou, P.; Chiu, C. Tuning the Circular Dichroism and Circular Polarized Luminescence Intensities of Chiral 2D Hybrid Organic–Inorganic Perovskites through Halogenation of the Organic Ions. *Angew. Chem. Int. Ed.* **2021**, *60* (39), 21434–21440. https://doi.org/10.1002/anie.202107239.

(16) Ahn, J.; Ma, S.; Kim, J.-Y.; Kyhm, J.; Yang, W.; Lim, J. A.; Kotov, N. A.; Moon, J. Chiral 2D Organic Inorganic Hybrid Perovskite with Circular Dichroism Tunable Over Wide Wavelength Range. *J. Am. Chem. Soc.* **2020**, *142* (9), 4206–4212. https://doi.org/10.1021/jacs.9b11453.

(17) Scalon, L.; Vaynzof, Y.; Nogueira, A. F.; Oliveira, C. C. How Organic Chemistry Can Affect Perovskite Photovoltaics. *Cell Rep. Phys. Sci.* **2023**, *4* (5), 101358. https://doi.org/10.1016/j.xcrp.2023.101358.

(18) Long, G.; Jiang, C.; Sabatini, R.; Yang, Z.; Wei, M.; Quan, L. N.; Liang, Q.; Rasmita, A.; Askerka, M.; Walters, G.; Gong, X.; Xing, J.; Wen, X.; Quintero-Bermudez, R.; Yuan, H.; Xing, G.; Wang, X. R.; Song, D.; Voznyy, O.; Zhang, M.; Hoogland, S.; Gao, W.; Xiong, Q.; Sargent, E. H. Spin Control in Reduced-Dimensional Chiral Perovskites. *Nat. Photonics* **2018**, *12* (9), 528–533. https://doi.org/10.1038/s41566-018-0220-6.

(19) Pietropaolo, A.; Mattoni, A.; Pica, G.; Fortino, M.; Schifino, G.; Grancini, G. Rationalizing the Design and Implementation of Chiral Hybrid Perovskites. *Chem* **2022**, *8* (5), 1231–1253. https://doi.org/10.1016/j.chempr.2022.01.014.

(20) Wang, Z.; Lin, C.; Murata, K.; Kamal, A. S. A.; Lin, B.; Chen, M.; Tang, S.; Ho, Y.; Chen, C.; Chen, C.; Daiguji, H.; Ishii, K.; Delaunay, J. Chiroptical Response Inversion and Enhancement of Room-Temperature Exciton-Polaritons Using 2D Chirality in Perovskites. *Adv. Mater.* **2023**, *35* (42), 2303203. https://doi.org/10.1002/adma.202303203.

(21) Ma, J.; Wang, H.; Li, D. Recent Progress of Chiral Perovskites: Materials, Synthesis, and Properties. *Adv. Mater.* **2021**, *33* (26), 2008785. https://doi.org/10.1002/adma.202008785.

(22) Jana, M. K.; Song, R.; Liu, H.; Khanal, D. R.; Janke, S. M.; Zhao, R.; Liu, C.; Valy Vardeny, Z.; Blum, V.; Mitzi, D. B. Organic-to-Inorganic Structural Chirality Transfer in a 2D Hybrid Perovskite and Impact on Rashba-Dresselhaus Spin-Orbit Coupling. *Nat. Commun.* **2020**, *11* (1), 4699. https://doi.org/10.1038/s41467-020-18485-7.

(23) Scalon, L.; Brunner, J.; Guaita, M. G. D.; Szostak, R.; Albaladejo-Siguan, M.; Kodalle, T.; Guerrero-León, L. A.; Sutter-Fella, C.

(23) M.; Oliveira, C. C.; Vaynzof, Y.; Nogueira, A. F. Tuning Phase Purity in Chiral 2D Perovskites. *Adv. Opt. Mater.* **2024**, *12* (2), 2300776. https://doi.org/10.1002/adom.202300776.

(24) Tanaka, K.; Takahashi, T.; Kondo, T.; Umeda, K.; Ema, K.; Umebayashi, T.; Asai, K.; Uchida, K.; Miura, N. Electronic and Excitonic Structures of Inorganic–Organic Perovskite-Type Quantum-Well Crystal ($C_4H_9NH_3$)$_2$PbBr$_4$. *Jpn. J. Appl. Phys.* **2005**, *44* (8R), 5923. https://doi.org/10.1143/JJAP.44.5923.

(25) Yaffe, O.; Chernikov, A.; Norman, Z. M.; Zhong, Y.; Velauthapillai, A.; Van Der Zande, A.; Owen, J. S.; Heinz, T. F. Excitons in Ultrathin Organic-Inorganic Perovskite Crystals. *Phys. Rev. B* **2015**, *92* (4), 045414. https://doi.org/10.1103/PhysRevB.92.045414.

(26) Mitzi, D. B.; Chondroudis, K.; Kagan, C. R. Organic-Inorganic Electronics. *IBM J. Res. Dev.* **2001**, *45* (1), 29–45. https://doi.org/10.1147/rd.451.0029.

(27) Haug, H.; Koch, S. W. *Quantum Theory of the Optical and Electronic Properties of Semiconductors*, 5th ed.; WORLD SCIENTIFIC, 2009. https://doi.org/10.1142/7184.

(28) Seitz, M.; Magdaleno, A. J.; Alcázar-Cano, N.; Meléndez, M.; Lubbers, T. J.; Walraven, S. W.; Pakdel, S.; Prada, E.; Delgado-Buscalioni, R.; Prins, F. Exciton Diffusion in Two-Dimensional Metal-Halide Perovskites. *Nat. Commun.* **2020**, *11* (1), 2035. https://doi.org/10.1038/s41467-020-15882-w.

(29) Deng, S.; Shi, E.; Yuan, L.; Jin, L.; Dou, L.; Huang, L. Long-Range Exciton Transport and Slow Annihilation in Two-Dimensional Hybrid Perovskites. *Nat. Commun.* **2020**, *11* (1), 664. https://doi.org/10.1038/s41467-020-14403-z.

(30) Billing, D. G.; Lemmerer, A. Synthesis and Crystal Structures of Inorganic–Organic Hybrids Incorporating an Aromatic Amine with a Chiral Functional Group. *CrystEngComm* **2006**, *8* (9), 686–695. https://doi.org/10.1039/B606987H.

(31) Seitz, M.; Gant, P.; Castellanos-Gomez, A.; Prins, F. Long-Term Stabilization of Two-Dimensional Perovskites by Encapsulation with Hexagonal Boron Nitride. *Nanomaterials* **2019**, *9* (8), 1120. https://doi.org/10.3390/nano9081120.

(32) Wu, X.; Trinh, M. T.; Niesner, D.; Zhu, H.; Norman, Z.; Owen, J. S.; Yaffe, O.; Kudisch, B. J.; Zhu, X.-Y. Trap States in Lead Iodide Perovskites. *J. Am. Chem. Soc.* **2015**, *137* (5), 2089–2096. https://doi.org/10.1021/ja512833n.

(33) Ma, J.; Fang, C.; Chen, C.; Jin, L.; Wang, J.; Wang, S.; Tang, J.; Li, D. Chiral 2D Perovskites with a High Degree of Circularly Polarized Photoluminescence. *ACS Nano* **2019**, *13* (3), 3659–3665. https://doi.org/10.1021/acsnano.9b00302.

(34) Scalon, L.; New, A.; Ge, Z.; Mondal, N.; Campos, R. D.; Quarti, C.; Beljonne, D.; Nogueira, A. F.; Bakulin, A. A.; Vaynzof, Y. Understanding and Controlling the Photoluminescence Line Shapes of 2D Perovskites with Chiral Methylbenzylammonium-Based Cations. *Chem. Mater.* **2024**, *36* (9), 4331–4342. https://doi.org/10.1021/acs.chemmater.3c03234.

(35) Li, X.; Hoffman, J. M.; Kanatzidis, M. G. The 2D Halide Perovskite Rulebook: How the Spacer Influences Everything from the Structure to Optoelectronic Device Efficiency. *Chem. Rev.* **2021**, *121* (4), 2230–2291. https://doi.org/10.1021/acs.chemrev.0c01006.

(36) Neumann, A.; Lindlau, J.; Colombier, L.; Nutz, M.; Najmaei, S.; Lou, J.; Mohite, A. D.; Yamaguchi, H.; Högele, A. Opto-Valleytronic Imaging of Atomically Thin Semiconductors. *Nat. Nanotechnol.* **2017**, *12* (4), 329–334. https://doi.org/10.1038/nnano.2016.282.

(37) Niu, L.; Liu, X.; Cong, C.; Wu, C.; Wu, D.; Chang, T. R.; Wang, H.; Zeng, Q.; Zhou, J.; Wang, X.; Fu, W.; Yu, P.; Fu, Q.; Najmaei, S.; Zhang, Z.; Yakobson, B. I.; Tay, B. K.; Zhou, W.; Jeng, H. T.; Lin, H.; Sum, T. C.; Jin, C.; He, H.; Yu, T.; Liu, Z. Controlled Synthesis of Organic/Inorganic van Der Waals Solid for Tunable Light–Matter Interactions. *Adv. Mater.* **2015**, *27* (47), 7800–7808. https://doi.org/10.1002/adma.201503367.

(38) Ginsberg, N. S.; Tisdale, W. A. Spatially Resolved Photogenerated Exciton and Charge Transport in Emerging Semiconductors. *Annu. Rev. Phys. Chem.* **2020**, *71* (1), 1–30. https://doi.org/10.1146/annurev-physchem-052516-050703.

(39) Akselrod, G. M.; Deotare, P. B.; Thompson, N. J.; Lee, J.; Tisdale, W. A.; Baldo, M. A.; Menon, V. M.; Bulović, V. Visualization of Exciton Transport in Ordered and Disordered Molecular Solids. *Nat. Commun.* **2014**, *5* (1), 3646. https://doi.org/10.1038/ncomms4646.

(40) Xiao, X.; Wu, M.; Ni, Z.; Xu, S.; Chen, S.; Hu, J.; Rudd, P. N.; You, W.; Huang, J. Ultrafast Exciton Transport with a Long Diffusion Length in Layered Perovskites with Organic Cation Functionalization. *Adv. Mater.* **2020**, *32* (46), 2004080. https://doi.org/10.1002/adma.202004080.

(41) Ziegler, J. D.; Zipfel, J.; Meisinger, B.; Menahem, M.; Zhu, X.; Taniguchi, T.; Watanabe, K.; Yaffe, O.; Egger, D. A.; Chernikov, A. Fast and Anomalous Exciton Diffusion in Two-Dimensional Hybrid Perovskites. *Nano Lett.* **2020**, *20* (9), 6674–6681. https://doi.org/10.1021/acs.nanolett.0c02472.

(42) Maiberg, M.; Hölscher, T.; Zahedi-Azad, S.; Scheer, R. Theoretical Study of Time-Resolved Luminescence in Semiconductors. III. Trap States in the Band Gap. *J. Appl. Phys.* **2015**, *118* (10), 105701. https://doi.org/10.1063/1.4929877.

(43) Kulig, M.; Zipfel, J.; Nagler, P.; Blanter, S.; Schüller, C.; Korn, T.; Paradiso, N.; Glazov, M. M.; Chernikov, A. Exciton Diffusion and Halo Effects in Monolayer Semiconductors. *Phys. Rev. Lett.* **2018**, *120* (20), 207401. https://doi.org/10.1103/PhysRevLett.120.207401.

(44) Mouri, S.; Miyauchi, Y.; Toh, M.; Zhao, W.; Eda, G.; Matsuda, K. Nonlinear Photoluminescence in Atomically Thin Layered WSe$_2$ Arising from Diffusion-Assisted Exciton-Exciton Annihilation. *Phys. Rev. B* **2014**, *90* (15), 155449. https://doi.org/10.1103/PhysRevB.90.155449.

(45) Ugur, E.; Ledinský, M.; Allen, T. G.; Holovský, J.; Vlk, A.; De Wolf, S. Life on the Urbach Edge. *J. Phys. Chem. Lett.* **2022**, *13* (33), 7702–7711. https://doi.org/10.1021/acs.jpclett.2c01812.

(46) Pan, Y.; Inam, F.; Zhang, M.; Drabold, D. A. Atomistic Origin of Urbach Tails in Amorphous Silicon. *Phys. Rev. Lett.* **2008**, *100* (20), 206403. https://doi.org/10.1103/PhysRevLett.100.206403.

(47) Ledinsky, M.; Schönfeldová, T.; Holovský, J.; Aydin, E.; Hájková, Z.; Landová, L.; Neyková, N.; Fejfar, A.; De Wolf, S. Temperature Dependence of the Urbach Energy in Lead Iodide Perovskites. *J. Phys. Chem. Lett.* **2019**, *10* (6), 1368–1373. https://doi.org/10.1021/acs.jpclett.9b00138.

(48) Caselli, V. M.; Wei, Z.; Ackermans, M. M.; Hutter, E. M.; Ehrler, B.; Savenije, T. J. Charge Carrier Dynamics upon Sub-Bandgap

Excitation in Methylammonium Lead Iodide Thin Films: Effects of Urbach Tail, Deep Defects, and Two-Photon Absorption. *ACS Energy Lett.* **2020**, *5* (12), 3821–3827. https://doi.org/10.1021/acsenergylett.0c02067.

(49) Zhang, Y.; Wang, R.; Li, Y.; Wang, Z.; Hu, S.; Yan, X.; Zhai, Y.; Zhang, C.; Sheng, C. Optical Properties of Two-Dimensional Perovskite Films of $(C_6H_5C_2H_4NH_3)_2[PbI_4]$ and $(C_6H_5C_2H_4NH_3)_2(CH_3NH_3)_2[Pb_3I_{10}]$. *J. Phys. Chem. Lett.* **2019**, *10* (1), 13–19. https://doi.org/10.1021/acs.jpclett.8b03458.